\documentstyle[fleqn,12pt]{article}

\begin{document}
\baselineskip=0.74cm
\begin{titlepage}
\begin{center}
\vspace{3.0in}
{\Large \bf Applications of Quantum Group to  Fractional Quantum Hall  Effect}
\end{center}

\begin{center}
{\bf    Guang-Hong Chen  and Mo-Lin Ge} \\
{\small {\it   Theoretical Physics Dvision, Nankai Institute of
Mathematics,}}\\
{\small {\it   Tianjin 300071, People's Republic of China }}\\
\end{center}
\vspace{2.0cm}
\begin{abstract}
\baselineskip=0.90cm
We show that there exists quantum group symmetry $ sl_{q}(2) $
in the fractional quantum Hall effect (FQHE) and  this
 symmetry  governs the degeneracy of ground-state level.
Under the  periodic boundary condition,
 the degree of  degeneracy   is related
  to the cyclic representation  of $sl_{q}(2)$.
   We also discuss the influence
  of  impurity by using  quantum group technique and give  the
    energy correction due to the impurity potential.

\noindent PACS number(s): 73.20H, 02.20.+b
\end{abstract}
\hspace{0.5cm}
\end{titlepage}
\newpage
\section{Introduction}
In the past ten years,  quantum groups(including Yangian [1] and quantum
algebras [2]) and their representation theories were intensively studied from
the
point of view of mathematical physics [3,4]. The simplest quantum algebra is
$sl_{q}(2)$ which can be viewed as a q-deformation of  classical Lie algebra
$sl(2)$
through the q-deformed commutation relations:
\begin{equation}
[J_{+}, J_{-}]=[2J_{3}]_{q}
\end{equation}
\begin{equation}
q^{J_{3}}J_{\pm}q^{-J_{3}}=q^{\pm 1}J_{\pm}
\end{equation}
and the related co-products, where we have used the notation:
\begin{equation}
[x]_{q}=\frac{q^{x}-q^{-x}}{q-q^{-1}}
\end{equation}
and $q$ stands for a complex deformation parameter. The representation for
Eqs.(1) and (2) are strongly dependent on whether $q$ at root of unity or
not. For the case where $q^{p}\ne 1$ ($p=2,3,...$), the representations are
similar to those of
$sl(2)$ algebra with the only difference  that  Clebsch-Gordan(C-G)
coefficients
should be replaced by the q-deformed ones and there are still the highest
weight and the lowest weight [5] as appeared in the usual Lie algebras.
However,
  it is not the case for $q^{p}=1$  whose representation does no longer
preserve
  the corresponding Lie algebraic  structure [6,7,8]. Or rather, it allows some
new
  types of representations   forbidden by  classical
Lie algebras. The typical one among them is the cyclic representation which
has neither the highest weight nor the lowest weight. In  physics  if a moving
particle experiences  an external magnetic field, then the deformation
parameter
$q$ is often related to the applied flux $\Phi$ through:
\begin{equation}
q={\rm e}^{i\Phi}
\end{equation}
The case with $q$ at root of unity means that the magnetic flux $\Phi$ is
quantized.

As is known that the quantum algebras shed a new light on the new symmetry to
the quantum
integrable systems in physics[9]. More attractively, they can be related to
some
interesting physical models. Among them  Wiegmann-Zabrodin [10],
Faddeev-Kashaev [11]
 and Hatsugai-Kohmoto-Wu [12] have accomplished their
remarkble works in this respect. In the Ref. [10], the Azbel-Hofstadter-Wannier
 problem of two-dimension Bloch electrons in  magnetic field was
  rediscussed by using the techniques of quantum group. Furthermore, in
   Ref. [11] the idea was extended to
the Chiral Potts model. However, it is still very  intereting to find other
physical examples
 which can be described in terms of $sl_{q}(2)$, especially, its cyclic
representation.

In this paper,  we would like to apply the cyclic representation to  fractional
quantum Hall effect(FQHE) which has extensively attracted  attention of
physicists
 for many years [13,14].
An important feature of FQHE is the degeneracy of  ground state [15,16] and
the degeneracy can be removed by impurity [17,18]. In the present paper  we
shall
 show the following  points:

(1) The degeneracy of  quantum states for FQHE can be described in terms of the
cyclic
representations of $sl_{q}(2)$. We find a proper quantum number $k$ which can
labels the degenerate quantum states;

(2) The potential for weak impurity can be reduced to a special case of the
  model which has been detailly studied by L. D. Faddeev and R. M. Kashaev
 in Ref. [11] and  Bethe Ansatz in Ref.[11] can be used to calculate the energy
 correction due
 to the weak impurity potential.
This is made through translating the method given by Tao-Haldane [17] into
quantum
algebraic  language. However, it is of interest since the results of
Faddeev-Kashaev
is concerned with the anisotropic extension of Wiegmann's approach for
Hofstadter model
and so far it has not been related to any "real physics".

This paper is organized as  follows. For self-containing we first introduce the
cyclic representation of quantum algebra through a simple physical realization.
In Sec.3 we  show how to explicitly define the quantum group symmetry in
single particle system and many-body system. In sec.4 we give the relation
between
irreducible cyclic representation of quantum algebra and  degeneracy for ground
 state of FQHE.
 In Sec.5  the removing of
 degeneracy due to impurity
will be discussed.  The final concluding remarks  will be made in Sec.6.

\section{Cyclic Representation of Quantum Algebras}
A simple physical realization of cyclic representation can be made through
the following example. We go along the line  of Pegg-Barnett(PB) theory
 on the  quantization of phase [19]. One can consider a harmonic oscilator
  in finite dimension Hilbert space and the set
   ${{\left |n\right\rangle,(n=0,1,...,s)}}$
  is the basis of the space. $\hat{N}$ is the "number" operator in  above
Hilbert
  space.
Then one can define phase states  as:
\begin{equation}
\left|\theta_{m}\right\rangle=\frac{1}{\sqrt{s+1}}\sum_{n=0}^{s}{\rm
e}^{in\theta_{m}}
\left|n\right\rangle
\end{equation}
where
\begin{equation}
\theta_{m}=\theta_{0}+\frac{2m\pi}{s+1}
\end{equation}
and $\theta_{0}$ is an arbitrary constant.
It is easy to check that  orthogonality and completeness hold:
\begin{equation}
\langle \theta_{p}|\theta_{m}\rangle=\delta_{pm},\hspace{2.0cm}
\sum_{m=0}^{s}\left|\theta_{m}\right\rangle\langle \theta_{m}|=1
\end{equation}
The PB phase operator can be defined as[19]:
\begin{equation}
\hat{\Phi}_{\theta}=\sum_{m=0}^{s}\theta_{m}\left|\theta_{m}
\right\rangle\langle
\theta_{m}|
\end{equation}
which satisfies:
\begin{equation}
{\rm e}^{i\hat{\Phi}_{\theta}}\left|n\right\rangle=\left|n-1\right\rangle
\hspace{1.0cm}
(n\ne 0),    \hspace{2.0cm}
{\rm e}^{i\hat{\Phi}_{\theta}}\left|0\right\rangle={\rm
e}^{i(s+1)\theta_{0}}\left|s\right\rangle
\end{equation}

\begin{equation}
{\rm e}^{-i\hat{\Phi}_{\theta}}\left|n\right\rangle=\left|n+1\right\rangle
\hspace{1.0cm}
(n\ne s) ,   \hspace{2.0cm}
{\rm e}^{-i\hat{\Phi}_{\theta}}\left|s\right\rangle={\rm
e}^{-i(s+1)\theta_{0}}\left|0\right\rangle
\end{equation}

Denoting
\begin{equation}
q={\rm e}^{i\frac{2\pi}{s+1}} ,\hspace{2.0cm}  (q^{s+1}=1)
\end{equation}
and defining
\begin{equation}
q^{\hat{N}}\left |n\right\rangle=q^{n+\eta}\left |n\right\rangle \hspace{1.0cm}
(n=0,1,...,s)
\end{equation}
where $\eta$ stands for an arbitrary phase factor, one can prove that
 the following equations hold:
\begin{eqnarray}
q^{\hat{N}}{\rm e}^{i\hat{\Phi}_{\theta}}&=&q^{-1}{\rm
e}^{i\hat{\Phi}_{\theta}}q^{\hat{N}} \\
q^{\hat{N}}{\rm e}^{-i\hat{\Phi}_{\theta}}&=&q{\rm
e}^{-i\hat{\Phi}_{\theta}}q^{\hat{N}}
\end{eqnarray}

\begin{equation}
{\rm e}^{i\hat{\Phi}_{\theta}}{\rm e}^{-i\hat{\Phi}_{\theta}}={\rm
e}^{-i\hat{\Phi}_{\theta}}
{\rm e}^{i\hat{\Phi}_{\theta}}
\end{equation}

Therefore, if one introduces:
\begin{equation}
b^{\dag}b=[\hat{N}]=\frac{q^{\hat{N}}-q^{-\hat{N}}}{q-q^{-1}}  \\
\end{equation}

\begin{eqnarray}
{\rm e}^{i\hat{\Phi}_{\theta}}=b[\hat{N}]^{-\frac{1}{2}}, \hspace{1.0cm}
{\rm e}^{-i\hat{\Phi}_{\theta}}=[\hat{N}]^{-\frac{1}{2}}b^{\dag}
\end{eqnarray}

then by making use of
Eqs.(16) and (17) we find that Eqs.(13),(14) and (15)
can be recast into
\begin{equation}
bb^{\dag}-q^{\pm}b^{\dag}b=q^{\pm\hat{N}}
\end{equation}

\begin{equation}
q^{\pm\hat{N}}b^{\dag}=q^{\pm}b^{\dag}q^{\pm\hat{N}}
\end{equation}

\begin{equation}
q^{\pm\hat{N}}b=q^{\mp}bq^{\pm\hat{N}}
\end{equation}

 Eqs.(18), (19) and (20) were known as  the deformed boson commutators of
quantum algebra
 $sl_{q}(2)$ [20].

   Eqs.(9) and (10) will become [20]:
\begin{equation}
b\left|n\right\rangle=\sqrt{[n+\eta]}\left|n-1\right\rangle \hspace{0.5cm}
(n\ne 0),    \hspace{2.0cm}
b\left|0\right\rangle=\sqrt{[\eta]}{\rm
e}^{i(s+1)\theta_{0}}\left|s\right\rangle
\end{equation}

\begin{equation}
b^{\dag}\left|n\right\rangle=\sqrt{[n+\eta+1]}\left|n+1\right\rangle
\hspace{0.5cm}
(n\ne s),    \hspace{1.5cm}
b^{\dag}\left|s\right\rangle=\sqrt{[\eta]}{\rm
e}^{-i(s+1)\theta_{0}}\left|0\right\rangle
\end{equation}

where the notation $[x]=\frac{q^{x}-q^{-x}}{q-q^{-1}}$ has been used. Above
 properties of state vectors
 are called the cyclic representation of quantum algebra  $sl_{q}(2)$ and can
be
 illustrated by  Fig.1.
\begin{figure}
\begin{picture}(450,150)(0,0)
\put(50,50){\oval(50,20)}
\put(140,50){\oval(50,20)}
\put(230,50){\oval(50,20)}
\put(320,50){\oval(50,20)}
\put(410,50){\oval(50,20)}
\put(80,55){\vector(1,0){30}}
\put(170,55){\vector(1,0){30}}
\put(260,55){\vector(1,0){30}}
\put(350,55){\vector(1,0){30}}
\put(110,45){\vector(-1,0){30}}
\put(200,45){\vector(-1,0){30}}
\put(290,45){\vector(-1,0){30}}
\put(380,45){\vector(-1,0){30}}
\put(100,61){\makebox(0,0){$b^{\dag}$}}
\put(190,61){\makebox(0,0){$b^{\dag}$}}
\put(280,61){\makebox(0,0){$b^{\dag}$}}
\put(370,61){\makebox(0,0){$b^{\dag}$}}
\put(100,39){\makebox(0,0){$b$}}
\put(190,39){\makebox(0,0){$b$}}
\put(280,39){\makebox(0,0){$b$}}
\put(370,39){\makebox(0,0){$b$}}
\footnotesize
\put(50,50){\makebox(0,0){$\left |0\right\rangle $}}
\put(140,50){\makebox(0,0){$\left |1\right\rangle $}}
\put(230,50){\makebox(0,0){...}}
\put(320,50){\makebox(0,0){$\left |s-1\right\rangle $}}
\put(410,50){\makebox(0,0){$\left |s\right\rangle $}}
\put(230,60){\oval(360,20)[t]}
\put(230,40){\oval(360,20)[b]}
\put(230,76){\makebox(0,0){$b^{\dag}$}}
\put(230,23){\makebox(0,0){$b$}}
\put(240,70){\vector(-1,0){10}}
\put(220,30){\vector(1,0){10}}

\end{picture}
\caption{Schematic Diagram of Cyclic Representation}
\end{figure}
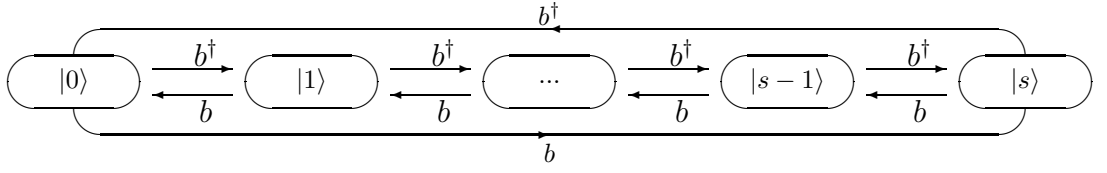
  There is another kind of realization of
Eqs.(13)-(15) that will be useful for the later dicussion. In order to make it
explicit
 one can define:
\begin{equation}
q^{\hat{N}}=X, \hspace{1.5cm}  {\rm e}^{i\hat{\Phi}_{\theta}}=Z
\end{equation}
Then Eq.(13) becomes
\begin{equation}
ZX=qXZ \hspace{2.0cm}  (q^{s+1}=1)
\end{equation}
which is Heisenberg-Weyl algebra. Based on it one is able to
construct the translation operators by setting:
\begin{eqnarray}
T_{x}=ZX , \hspace{1.5cm}   T_{y}=ZX^{-1}
\end{eqnarray}

It is easy to check:
\begin{eqnarray}
T_{y}T_{x}=q^{2}T_{x}T_{y}     \hspace{2.0cm}
 T_{-y}T_{-x}=q^{2}T_{-x}T_{-y}  \\
T_{-y}T_{x}=q^{-2}T_{x}T_{-y} \hspace{2.0cm}
 T_{y}T_{-x}=q^{-2}T_{-x}T_{y}
\end{eqnarray}
which is identical with  quantum algebra denoted by the commutation relations
 Eqs.(18), (19) and (20) through  defining:
\begin{equation}
T_{x}+T_{y}=i(q-q^{-1})J_{-}    \hspace{2.0cm}
T_{-x}+T_{-y}=i(q-q^{-1})J_{+}
\end{equation}

\begin{equation}
T_{-x}T_{y}=qK^{-2}                \hspace{2.5cm}
T_{-y}T_{x}=q^{-1}K^{+2}
\end{equation}

We then  have
\begin{equation}
[J_{+},J_{-}]=\frac{K^{+2}-K^{-2}}{q-q^{-1}}
\end{equation}

\begin{equation}
K^{+}J_{\pm}K^{-}=q^{\pm 1}J_{\pm }
\end{equation}
By straightforword calculation it is easy to find:
\begin{equation}
K^{+}=q^{\hat{N}+\frac{1}{2}} \hspace{2.0cm}
K^{-}=q^{-\hat{N}-\frac{1}{2}}
\end{equation}
and
\begin{equation}
K^{\pm}\left |n\right\rangle=q^{\pm (n+\eta+\frac{1}{2})}\left |n\right\rangle
\end{equation}

\begin{equation}
J_{+}|n\rangle=-\frac{\cos{[\gamma(n+\eta +1)]}}{\sin{\gamma}}|n+1\rangle
\hspace{1.0cm}(n\ne s)
\end{equation}

\begin{equation}
J_{+}|s\rangle=-\frac{\cos{(\gamma\eta)}}{\sin{\gamma}}{\rm
e}^{-i(s+1)\theta_0}|0\rangle
\end{equation}

\begin{equation}
J_{-}|n\rangle=-\frac{\cos{[\gamma(n+\eta)]}}{\sin{\gamma}}|n-1\rangle
\hspace{1.0cm} (n\ne 0)
\end{equation}

\begin{equation}
J_{-}|0\rangle=-\frac{\cos{\gamma \eta}}{\sin{\gamma}}{\rm
e}^{i(s+1)\theta_0}|s\rangle
\end{equation}
where the notations $\gamma=\frac{2\pi}{s+1}$ and $q={\rm e}^{i\gamma}$ have
been
used. This  realization
 of quantum algebra $sl_{q}(2)$  is very useful and will be used in the
  following discussion.

\section{Magnetic Translation  Invariance and Quantum Group Symmetry}

     For simplity  we first consider a spinless particle which moves in a plane
and
experiences an uniform external magnetic field along $z$-direction, $
\vec{B}=B\hat{e}_{z}$. The Hamiltanian of  system can be written
as
\begin{equation}
 H_{0}=\frac{1}{2m}(\vec{p}+e\vec{A})^2
\end{equation}
where $m$,$e$ are  mass and charge of  particle, respectively. $
\vec{A} $ is the  vector potential satisfying:
\begin{equation}
\bigtriangledown \times \vec{A}=B\hat{e}_{z}
\end{equation}

Above problem can  easily be solved in  proper gauge  [21,22].
We shall discuss the
 gauge-independent case.
 It is well known that in this  system there is not    translation invariance,
    however, it  exhibits
magnetic translation invariance  generated by the magnetic translation
operator  defined by
\begin{equation}
 t(\vec{a})=\exp
[\frac{i}{\hbar}\vec{a}\cdot(\vec{p}+e\vec{A}+e\vec{r}\times\vec{B})]
\end{equation}
where $\vec{a}=a_{x}\hat{e}_{x}+a_{y}\hat{e}_{y}$ is an arbitrary
two-dimensional
vector.
The magnetic translation operator $t(\vec{a})$ satisfies the following group
property [23,24]:
\begin{equation}
 t(\vec{a})t(\vec{b})=\exp
[-i\frac{\hat{e}_{z}\cdot(\vec{a}\times\vec{b})}{a_{0}^2}] t(\vec{b})t(\vec{a})
\end{equation}
where $a_{0}\equiv\sqrt{\frac{\hbar}{eB}}$ is the magnetic length.

Let
\begin{equation}
\vec{\kappa}=\vec{p}+e\vec{A}+e\vec{r}\times\vec{B}
\end{equation}

It is easy to prove that
\begin{equation}
 [t(\vec{a}),H_{0}]=0,\hspace{2 cm} [\vec{\kappa},H_{0}]=0
\end{equation}
i.e. the system under consideration is invariant under
the magnetic translation transformation, and $\vec{\kappa}$ is
a conservative quantity.

With the help of the magnetic translation operator, one can construct the
following operators [10]:
\begin{equation}
 J_{+}=\frac{1}{q-q^{-1}}[t(\vec{a})+t(\vec{b})],\hspace{1.0cm}
 J_{-}=\frac{-1}{q-q^{-1}}[t(-\vec{a})+t(-\vec{b})]
\end{equation}

\begin{equation}
 q^{2J_{3}}=t(\vec{b}-\vec{a}),\hspace{1.5cm}
 q^{-2J_{3}}=t(\vec{a}-\vec{b})
\end{equation}
with
\begin{equation}
 q=\exp (i2\pi\frac{\Phi}{\Phi_{0}})
\end{equation}
where $ \Phi=\frac{1}{2}\vec{B}\cdot(\vec{a}\times\vec{b}) $ is  magnetic
flux through the triangle  enclosed by vectors $\vec{a}$ and $\vec{b}$ and
$\Phi_{0}=\frac{h}{e}$ is  magnetic flux quanta.
It turns out  that the operators $J_{+}$,$J_{-}$ and $J_{3}$
satisfy the algebraic relations of the  $sl_{q}(2)$ [2] as shown by Eqs.(1) ,
(2)
and (46).

{}From Eqs.(43)-(45)  it follows that:
\begin{equation}
 [J_{\pm},H_{0}]=0
\end{equation}

\begin{equation}
 [q^{\pm J_{3}},H_{0}]=0
\end{equation}
which indicates that $J_{\pm}$ and $J_{3}$ are conservative quantities of the
system. Therefore, there is the  $sl_{q}(2)$ structure in the Landau problem
under our consideration. This structure still holds for
many-body system. We assume that a system contains  $N_{e}$ electrons and the
interaction
among  particles is pair-potential $V(\mid \vec{r_{i}}-\vec{r}_{j}\mid)$,
where $\vec{r}_{i}$ is the coordinate of the $i-th$ electron.
 The Hamiltonian of  system in the absence of  impurities can be written as:
 \begin{equation}
 H=\sum_{j=1}^{N_{e}}\frac{1}{2m}(\vec{p}_{j}+e\vec{A_{j}})^{2}+\sum_{i\ne
j}^{N_{e}}V(\mid\vec{r}_{i}-\vec{r}_{j}\mid)
 \end{equation}
 In this case one can construct the following generators of magnetic
translation:
 \begin{equation}
 T(\vec{a})=\prod_{j=1}^{N_{e}}t_{j}(\vec{a})
 =\exp (\frac{i}{\hbar}N_{e}\vec{\kappa}_{c}\cdot\vec{a})
 \end{equation}
 where
 \begin{equation}
 \vec{\kappa}_{c}=\frac{1}{N_{e}}\sum_{j=1}^{N_{e}}\vec{\kappa}_{j}
 \end{equation}
 is the psedo-momentum of mass centre, $N_{e}$ stands for the number of
  electrons  and $\vec{\kappa}_{j}$ is the psedo-momentum of the $j-th$
   electron.
 It is easy to know:
 \begin{equation}
T_{\vec{a}}T_{\vec{b}}=\exp
[-iN_{e}\frac{\hat{e}_{z}\cdot(\vec{a}\times\vec{b})}
{a_{0}^2}]T_{\vec{b}}T_{\vec{a}}
\end{equation}
Similarly, by defining
\begin{equation}
 J_{+}=\frac{1}{q-q^{-1}}[T(\vec{a})+T(\vec{b})],\hspace{1.0cm}
 J_{-}=\frac{-1}{q-q^{-1}}[T(-\vec{a})+T(-\vec{b})]
\end{equation}

\begin{equation}
 q^{2J_{3}}=T(\vec{b}-\vec{a}),\hspace{1.5cm}
 q^{-2J_{3}}=T(\vec{a}-\vec{b})
\end{equation}
with
\begin{equation}
 q=\exp [iN_{e}\frac{\hat{e}_{z}\cdot(\vec{a}\times\vec{b})}{2a_{0}^2}]
\end{equation}
We still obtain
\begin{eqnarray}
 [J_{+},J_{-}]&=&[2J_{3}]_{q}   \nonumber    \\
 q^{J_{3}}J_{\pm}q^{-J_{3}}&=&q^{\pm 1}J_{\pm}
\end{eqnarray}
In other words, we can also define a  quantum group describing the global
behavior of a many-body system.

Because $T(\vec{a})$ is the magnetic translation operator of  mass centre,
 it can not change the relative distance among  particles, namely,
 \begin{equation}
[T(\vec{a}),V(\mid \vec{r_{i}}-\vec{r}_{j}\mid)]=0
\end{equation}
Similarly, one can also  check
\begin{equation}
[T(\vec{a}),\sum_{j=1}^{N_{e}}\frac{1}{2m}(\vec{p}_{j}+e\vec{A_{j}})^{2}]=0
\end{equation}
We then obtain
\begin{equation}
[T(\vec{a}),H]=0
\end{equation}
Eqs.(53), (54) and (59) implies that:
\begin{equation}
 [J_{\pm},H]=0,\hspace{2 cm} [q^{\pm J_{3}},H]=0
\end{equation}
Therefore, the interacting electrons in  magnetic field
 also exhibit a hidden symmetry---quantum algebra $sl_{q}(2)$. The
similar result has been found in  interacting anyon system [25]. However, after
review the above results one
would like to ask what kind of effect can be caused by above symmetry. The
answer
will be given in  following section.

\section{ Cyclic Representation and Degeneracy of  Ground State for  FQHE}

As is known that the fractional quantum Hall system can be described by the
Hamiltanian
Eq.(49) and the pair-potential $V(|\vec{r}_{i}-\vec{r}_{j}|)$ is just the
Coloumb
interaction $\frac{e^{2}}{|\vec{r}_{i}-\vec{r}_{j}|}$.
Let $\Psi$  be  wave function of  system in the Schr\"{o}dinger picture.
Following  the basic idea of Wen and Niu [18],   we  employ
the periodic boundary condition (PBC):
\begin{equation}
 t_{j}(\vec{L}_{1})\Psi =\Psi,\hspace{2 cm}  t_{j}(\vec{L}_{2})\Psi=\Psi
\end{equation}
where $\vec{L}_{1}=L_{1}\hat{e}_{x}$,  $\vec{L}_{2}=L_{2}\hat{e}_{y}$, and
 $j=1,2,...,N_{e}$. This
boundary condition means that  particles are confined in a rectangular area of
 size $L_{1}\times L_{2}$.
{}From Eq.(61) it follows that the operators $t_{j}(\vec{L}_{1})$ and
$t_{j}(\vec{L}_{2})$
commute with each other. That is,
\begin{equation}
 t_{j}(\vec{L}_{1})t_{j}(\vec{L}_{2})=t_{j}(\vec{L}_{2})t_{j}(\vec{L}_{1})
\end{equation}
however, from Eq.(41) we have
\begin{equation}
t_{j}(\vec{L}_{1})t_{j}(\vec{L}_{2})=exp[-i\frac{\hat{e}_{z} \cdot(\vec{L}_{1}
\times\vec{L}_{2})}{a_{0}^{2}}]t_{j}(\vec{L}_{2})t_{j}(\vec{L_{1}})
\end{equation}
Combining Eq.(62) with Eq.(63) yields:
\begin{equation}
\exp (i2\pi\frac{\Phi}{\Phi_{0}})=1
\end{equation}
where $\Phi=\frac{1}{2}BL_{1}L_{2}$ is  magnetic flux through the triangle
enclosed
 by $\vec{L}_{1}$ and $\vec{L}_{2}$.  Eq.(64) implies that
\begin{equation}
\Phi=N_{s}\Phi_{0}
\end{equation}
where $N_{s}$ is a positive integer.
Therefore, the periodic boundary condition (61) is equivalent to   magnetic
 flux quantinization. It is well known that
  the Landau filling factor  $\nu$ satisfies
  \begin {equation}
  \nu=\frac{N_{e}}{N_{s}}=\frac{P}{Q}
  \end{equation}
   where $P$ and $Q$ are two mutal prime integers. One should
notice that when  the boundary condition Eq.(61) is taken, not all the
 translation operators $T(\vec{a})$ leaves  Eq.(61)
 invariant. In other words,
\begin{equation}
t_{j}(\vec{L}_{i})T(\vec{a})\Psi=T(\vec{a})\Psi   \nonumber \\
( i=1,2 )
\end{equation}
 can not be satisfied by an arbitrary maganetic translation $T(\vec{a})$.
However, if we define two primitive magnetic translation operators in the
following way [18]:
\begin{equation}
 T_{x}\equiv T(\frac{\vec{L}_{1}}{N_{s}}),\hspace{2 cm} T_{y}\equiv
T(\frac{\vec{L}_{2}}{N_{s}})
\end{equation}
then only $T_{x},T_{y}$ and their integer powers can make Eq.(67) hold.

By a straightforward calculation  it can be checked that the following
relations hold
\begin{equation}
T_{y}T_{x}=exp(i2\pi\frac{P}{Q})T_{x}T_{y},\hspace{1.0cm}
 T_{-y}T_{-x}=exp(i2\pi\frac{P}{Q})T_{-x}T_{-y}
\end{equation}

\begin{equation}
 T_{-y}T_{x}=exp(-i2\pi\frac{P}{Q})T_{x}T_{-y},\hspace{1.0cm}
 T_{y}T_{-x}=exp(-i2\pi\frac{P}{Q})T_{-x}T_{y}
\end{equation}

\begin{equation}
 T_{-x}T_{x}=T_{-y}T_{y}=1
\end{equation}
It is easy to see that the generators $T_{x},T_{y}$ are not enough to describe
the
specified phyical problem. However, similar to Ref. [10] by making use of the
 operators $T_{\pm{x}},
 T_{\pm{y}}$ and the  above commutation relations we can construct the
generators
 of
  quantum group  as follows:
\begin{equation}
 \hat{J}_+=\frac{-i}{q-q^{-1}}(T_{-x}+T_{-y}),\hspace{2
cm}\hat{J}_-=\frac{-i}{q-q^{-1}}(T_{x}+T_{y})
\end{equation}
\begin{equation}
  \hat{K}^{+2}=qT_{-y}T_{x},\hspace{2cm}  \hat{K}^{-2}=q^{-1}T_{-x}T_{y}
\end{equation}
where the deformation paramerter is given by
\begin{equation}
 q=\exp (i\pi\frac{P}{Q})
\end{equation}
It is easy to check that these generators obey the standard commutation
relations
of the quantum group $sl_{q}(2)$ as shown by Eqs.(30), (31) and (74).
We can also find that the generators $\hat{J}_{\pm}$ and $\hat{K}^{\pm}$
 commute with  Hamiltonian
\begin{equation}
 [\hat{J}_{\pm},H]=0,\hspace{2 cm} [\hat{K}^{\pm },H]=0
\end{equation}
Above analysis indicates that  $sl_{q}(2)$  is the basic symmetry in our
system.
Furthermore, according to the fundamental  principle of quantum mechanics,
Eqs.(30), (31) and (75)
  imply the  degeneracy of  ground state in the FQHE.

  Let us discuss
 the relationship between degeneracy  and  cyclic representation
 of $sl_{q}(2)$.
 Since $P,Q$ are integers  in Eq.(74),  we  should discuss the following two
cases:

 (i)\hspace{0.3cm}$ P=even $

 We have:
 \begin{equation}
 q^{Q}=1
 \end{equation}

   In this case  the representation of quantum
 group     has
  so-called cyclic representation and the dimension of the irreducible
   representation is $Q$ [7].

Furthermore, without loss of  generality, according to Eq.(75) we can
simutaneously
 diagonalize $H$ and $\hat{K}^{\pm}$. In other words, one can choose a set of
  basis vectors $\left | n,k\right\rangle=\left |n\right\rangle\otimes\left
|k\right\rangle $
   to be  eigenvectors of operators  $H$ and $\hat{K}^{\pm}$.
i.e.,
\begin{equation}
 H\left |n,k\right\rangle=E_{n}\left|n,k\right\rangle
\end{equation}
and
\begin{equation}
  \hat{K}^{\pm}\left |n,k\right\rangle=q^{\pm (k+\eta+\frac{1}{2})}\left
|n,k\right\rangle
\end{equation}
where $n=0,1,...,\infty$ is the symbols of the energy level, and
 $k=0,1,...,Q-1$ is the new quantum numbers which label  the
 different quantum states in the same degenerate energy level.
 The cyclic representation for this case is shown by Fig.2.
\begin{figure}
\begin{picture}(450,150)(0,0)
\put(50,50){\oval(50,20)}
\put(140,50){\oval(50,20)}
\put(230,50){\oval(50,20)}
\put(320,50){\oval(50,20)}
\put(410,50){\oval(50,20)}
\put(80,55){\vector(1,0){30}}
\put(170,55){\vector(1,0){30}}
\put(260,55){\vector(1,0){30}}
\put(350,55){\vector(1,0){30}}
\put(110,45){\vector(-1,0){30}}
\put(200,45){\vector(-1,0){30}}
\put(290,45){\vector(-1,0){30}}
\put(380,45){\vector(-1,0){30}}
\put(100,62){\makebox(0,0){$\hat{J}_+$}}
\put(190,62){\makebox(0,0){$\hat{J}_+$}}
\put(280,62){\makebox(0,0){$\hat{J}_+$}}
\put(370,62){\makebox(0,0){$\hat{J}_+$}}
\put(100,38){\makebox(0,0){$\hat{J}_-$}}
\put(190,38){\makebox(0,0){$\hat{J}_-$}}
\put(280,38){\makebox(0,0){$\hat{J}_-$}}
\put(370,38){\makebox(0,0){$\hat{J}_-$}}
\footnotesize
\put(50,50){\makebox(0,0){$\left |n,0\right\rangle $}}
\put(140,50){\makebox(0,0){$\left |n,1\right\rangle $}}
\put(230,50){\makebox(0,0){...}}
\put(320,50){\makebox(0,0){$\left |n,Q-2\right\rangle $}}
\put(410,50){\makebox(0,0){$\left |n,Q-1\right\rangle $}}
\put(230,60){\oval(360,20)[t]}
\put(230,40){\oval(360,20)[b]}
\put(230,77){\makebox(0,0){$\hat{J}_+$}}
\put(230,22){\makebox(0,0){$\hat{J}_-$}}
\put(240,70){\vector(-1,0){10}}
\put(220,30){\vector(1,0){10}}

\end{picture}
\caption{Schematic Diagram of Cyclic Representation}
\end{figure}
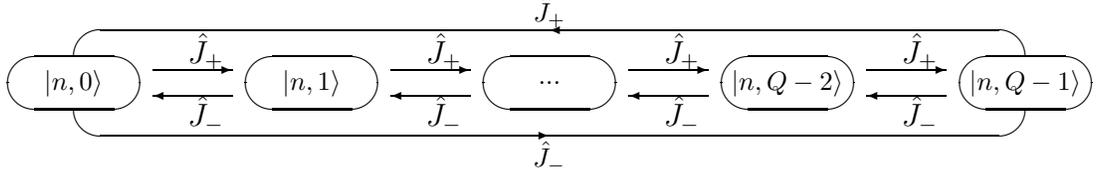
  According to Eqs.(33)-(37) and
 acting the $sl_{q}(2)$ generators on these basis vectors yield:
\begin{equation}
\hat{J}_-|n,k\rangle=-\frac{\cos{[\gamma(n+\eta)]}}{\sin{\gamma}}|n,k-1\rangle
\hspace{1.0cm}(k\ne 0)
\end{equation}

\begin{equation}
\hat{J}_-|n,0\rangle=-\frac{\cos{(\gamma\eta)}}{\sin{\gamma}}{\rm
e}^{iQ\theta_{0}}|n,s\rangle
\end{equation}

\begin{equation}
\hat{J}_+|n,k\rangle=-\frac{\cos{[\gamma(k+\eta+1)]}}{\sin{\gamma}}|n,k+1
\rangle \hspace{1.0cm} (k\ne Q-1)
\end{equation}

\begin{equation}
\hat{J}_+|n,s\rangle=-\frac{\cos{\gamma \eta}}{\sin{\gamma}}{\rm
e}^{-iQ\theta_{0}}|n,0\rangle
\end{equation}
where $\eta,\theta_{0} $ are arbitrary constants,  $q={\rm e}^{i\gamma}$
and  $\gamma=\pi\frac{P}{Q}$.

    Since the dimension of the irreducible cyclic representation space ${\left
|n,k\right\rangle }$
is $Q$, from Eq.(79)-(82) we can see that the degree of degeneracy of the
ground state
is just $Q$, which is in accordance with the statement of the current
literatures [15-18]. This is one of the main conclusions of this paper.

(ii)\hspace{0.3cm}$ P=odd $

This case is different from case (i). We have:
 \begin{equation}
 q^{2Q}=1
 \end{equation}
 Similar to case $(i)$  we can draw a similar conclusion about the degeneracy,
  but the degree of degeneracy is $2Q$. The properties of transformation among
  the degenerate states are similar to case $(i)$ except that $Q$ should be
  replaced by  $2Q$.

  Briefly speaking,  the degree of degeneracy of ground state depends on the
Landau
  filling factor $ \nu=\frac{P}{Q}$. There is $Q$-fold
  degeneracy for the ground state  for $P=even$, whereas  $2Q$-fold
  degeneracy for $P=odd$. The latter is somewhat different from the previous
   discussion [17,18,26].
   Our results concerning the degeneracy of energy
  are independent of the ground state or the excited states of the system. It
allows
  the mixture of different levels. Besides, we do not know how to eliminate
   the case (ii) from the point of view of symmetry-breaking.


\section{Influence of Impurity on the Degeneracy}

Although FQHE  can often be observed in the high mobility sample, the impurity
can not be avoided at any case.  It is interesting to inverstigate the
influence
of  impurity potential. We shall concentrate on the behavior of weak
impurity. In this section we first derive the effective impurity potential in
terms
of the generators of quantum group $sl_{q}(2)$, then we shall explore how the
degeneracy
of ground state is removed by means of   perturbation
theory. A  generalized Bethe-Ansatz equation will be used for deriving the
energy correction.

Generally speaking, the impurity potential can be written as [17]:
\begin{equation}
U=\sum_{j=1}^{N_{e}}V(r_{j})
=\sum_{\vec{k}}\tilde{V}(\vec{k})\sum_{j=1}^{N_{e}}{\rm
e}^{i\vec{k}\cdot\vec{r}_{j}}
\end{equation}
where $\tilde{V}(\vec{k})$ is the Fourier transform of $V(r_{j})$ and $\vec{k}$
 is the wave vector of Fourier transform which is defined as:
 \begin{equation}
k_{x}=\frac{2\pi}{L_{1}}n_{1}  ,\hspace{4.0cm}
k_{y}=\frac{2\pi}{L_{2}}n_{2}
\end{equation}
where $n_{1}$ and $n_{2}$ are integers.

Making use of the well-known  formula:
\begin{equation}
{\rm e}^{\hat{A}+\hat{B}}={\rm e}^{\hat{A}}{\rm e}^{\hat{B}}
{\rm e}^{-\frac{1}{2}[\hat{A},\hat{B}]}
\end{equation}
where $\hat{A}$ and $\hat{B}$ are two operators with $[\hat{A},\hat{B}]$
being constant, we  obtain:
\begin{equation}
{\rm e}^{i\vec{k}\cdot\vec{r}_{j}}={\rm e}^{-\frac{1}{2}k^{2}a_{0}^{2}}{\rm
e}^{\frac{k_{-}\Pi_{j+}}{2eB}}
{\rm e}^{i\frac{k_{y}\kappa_{jx}-k_{x}\kappa_{jy}}{eB}}{\rm
e}^{\frac{-k_{+}\Pi_{j-}}{2eB}}
\end{equation}
where $k_{\pm}$ and $\Pi_{j\pm}$ are defined by:
\begin{eqnarray}
k_{\pm}=k_{x}\pm ik_{y} , \hspace{2.0cm}
\Pi_{j\pm}=\Pi_{jx}\pm i\Pi_{jy}
\end{eqnarray}
The  operators  $\Pi_{j\pm}$ can  raise or  lower the Landau levels [17].
 Therefore, if  the influence  of impurity on the  ground state is mainly taken
into account,
  we only need to discuss the projection of $V(\vec{r}_{i})$  at the lowest
Landau
  level, that is:
\begin{equation}
U_{0}=\sum_{\vec{k}}\tilde{V}(\vec{k}){\rm
e}^{-\frac{1}{2}k^{2}a_{0}^{2}}\sum_{j=1}^{N_{e}}
{\rm e}^{i\frac{k_{y}\kappa_{jx}-k_{x}\kappa_{jy}}{eB}}
\end{equation}

Besides, from Eqs.(69) and (70) we know that ${T_{\pm x}^{Q},T_{\pm y}^{Q}}$
are the
centre elements of the $sl_{q}(2)$.
 They can then be taken as constants, i.e.,
 \begin{equation}
 T_{i}^{Q}{\rm e}^{i\vec{k}\cdot\vec{r}}={\rm
e}^{i\vec{k}\cdot\vec{r}}T_{i}^{Q}
\end{equation}
where $i=\pm{x}, \pm{y}$. However, through the detail caculation one can find
that:
 \begin{eqnarray}
 T_{\pm{x}}^{Q}{\rm e}^{i\vec{k}\cdot\vec{r}}={\rm
e}^{i\vec{k}\cdot\vec{r}}T_{\pm{x}}^{Q}{\rm e}^{\pm
i2\pi\frac{n_{1}Q}{N_{s}}}\nonumber\\
 T_{\pm{y}}^{Q}{\rm e}^{i\vec{k}\cdot\vec{r}}={\rm
e}^{i\vec{k}\cdot\vec{r}}T_{\pm{y}}^{Q}{\rm e}^{\pm i2\pi\frac{n_{2}Q}{N_{s}}}
\end{eqnarray}
In comparison to Eq.(90) one  obtains:
 \begin{eqnarray}
n_{1}=\frac{l_{1}N_{s}}{Q}  ,   \hspace{2.0cm}
n_{2}=\frac{l_{2}N_{s}}{Q}
\end{eqnarray}
where $l_{1}$ and $l_{2}$ are other new integers.

Substituting Eqs.(85) and (92) into Eq.(89) one has:
\begin{equation}
U_{0}=N_{e}\sum_{l_{1},l_{2}}\tilde{V}(l_{1}m_{1},l_{2}m_{2})
{\rm e}^{-\frac{1}{2}[(\frac{l_{1}L_{2}}{Qa_{0}})^{2}+(\frac{l_{2}L_{1}}
{Qa_{0}})^{2}]}t(\frac{l_{2}}{Q}\vec{L}_{1}-\frac{l_{1}}{Q}\vec{L}_{2})
\end{equation}
where  constants $m_{1}$ and $m_{2}$ are  defined as:
\begin{eqnarray}
m_{1}=\frac{L_{2}}{Qa_{0}^{2}}  , \hspace{2.0cm}
m_{2}=\frac{L_{1}}{Qa_{0}^{2}}
\end{eqnarray}
 and the magnetic translation operator
  $t(\frac{l_{2}}{Q}\vec{L}_{1}-\frac{l_{1}}{Q}\vec{L}_{2})$
  act on  electrons only.

  As is known that the Gaussian factor appeared in  Eq.(93) will make  $U_{0}$
  rapidly decay, therefore we can only take the leading term in the expansion
  of  $U_{0}$, i.e., to the
  lowest order of ${\rm e}^{-(\frac{L_{1}}{Qa_{0}})^{2}}$ and ${\rm
e}^{-(\frac{L_{2}}{Qa_{0}})^{2}}$
  ,  we  the  get:
  \begin{equation}
  U_{0}\doteq u_{1}t(\frac{\vec{L}_{1}}{Q})+u_{2}t(\frac{\vec{L}_{2}}{Q})+ h.c.
\end{equation}
where we have neglected the constant term and defined:
\begin{eqnarray}
u_{1}&=&N_{e}\tilde{V}(0,m_{2}){\rm
e}^{-\frac{1}{2}(\frac{L_{1}}{Qa_{0}})^{2}}\nonumber\\
u_{2}&=&N_{e}\tilde{V}(-m_{1},0){\rm
e}^{-\frac{1}{2}(\frac{L_{2}}{Qa_{0}})^{2}}
\end{eqnarray}
we should bear in mind that the set ${T_{\pm{x}}}$, and $T_{\pm{y}}$ is
complete
and closed for  $sl_{q}(2)$.
By tedious calculation
one  finds:
\begin{eqnarray}
[T_{x}^{-r}t(\frac{\vec{L}_{1}}{Q}), T_{x}]=0,
\hspace{2.0cm}[T_{y}^{-r}t(\frac{\vec{L}_{2}}{Q}), T_{y}]=0
\end{eqnarray}
and
\begin{eqnarray}
T_{x}^{-r}t(\frac{\vec{L}_{1}}{Q})T_{y}&=&T_{y}T_{x}^{-r}t(\frac{\vec{L}_{1}}
{Q})
{\rm e}^{i2\pi\frac{Pr-1}{Q}}\nonumber\\
T_{y}^{-r}t(\frac{\vec{L}_{2}}{Q})T_{x}&=&T_{x}T_{y}^{-r}t(\frac{\vec{L}_{2}}
{Q})
{\rm e}^{i2\pi\frac{Pr-1}{Q}}\nonumber\\
T_{x}^{-r}t(\frac{\vec{L}_{1}}{Q})T_{-y}&=&T_{-y}T_{x}^{-r}t(\frac{\vec{L}_{1}}
{Q})
{\rm e}^{-i2\pi\frac{Pr-1}{Q}}\nonumber\\
T_{y}^{-r}t(\frac{\vec{L}_{2}}{Q})T_{-x}&=&T_{-x}T_{y}^{-r}t(\frac{\vec{L}_{2}}
{Q})
{\rm e}^{-i2\pi\frac{Pr-1}{Q}}
\end{eqnarray}
Therefore, if we choose
\begin{equation}
Pr+Qs=1
\end{equation}
where $s$ is an integer, then the operator $T_{x}^{-r}t(\frac{\vec{L}_{1}}{Q})$
can commute with any operator of the set ${T_{\pm{x}}}$ and $T_{\pm{y}}$. The
same results
can be derived for the operator $T_{y}^{-r}t(\frac{\vec{L}_{2}}{Q})$.
 Taking Shur's lemma into account  the operators
$T_{x}^{-r}t(\frac{\vec{L}_{1}}{Q})$
 and $T_{y}^{-r}t(\frac{\vec{L}_{2}}{Q})$ are nothing but  constants. Without
loss of  generality we can choose:
\begin{eqnarray}
T_{x}^{-r}t(\frac{\vec{L}_{1}}{Q})={\rm e}^{i\phi_{1}}   \hspace{2.0cm}
T_{y}^{-r}t(\frac{\vec{L}_{2}}{Q})={\rm e}^{i\phi_{2}}
\end{eqnarray}
i.e., we have
\begin{eqnarray}
t(\frac{\vec{L}_{1}}{Q})={\rm e}^{i\phi_{1}}T_{x}^{r}    \hspace{2.0cm}
t(\frac{\vec{L}_{2}}{Q})={\rm e}^{i\phi_{2}}T_{y}^{r}
\end{eqnarray}
where $\phi_{1}$ and $\phi_{2}$ are two constants.
Furthermore, we can prove based on the number theory that the solution of
Eq.(99) is unique if we impose a
constraint on the value of $r$
\begin{equation}
\mid r \mid<\frac{Q}{2}
\end{equation}

That is to say, we can uniquely express $t(\frac{\vec{L}_{1}}{Q})$ and
$t(\frac{\vec{L}_{2}}{Q})$
in terms of $T_{x}^{r}$  and $T_{y}^{r}$. This is an interesting  result.
 Substituting Eq.(101) into Eq.(95) we obtain:
\begin{equation}
  U_{0}=u_{1}{\rm e}^{i\phi_{1}}T_{x}^{r}+u_{2}{\rm e}^{i\phi_{2}}T_{y}^{r}+
c.c.
\end{equation}
Because the impurity potential $U_{0}$ is only related to  operators $T_{x}$
and
$T_{y}$, but  not to $t_{j}(\vec{a})$, so we have to find the effective change
of  boundary conditions, or equivalently, we should find an effective operator
$U_{0}$ which  leaves the wave function unchanged. A simple form of the
operator
 can be
chosen as [18]:
\begin{equation}
  U_{0}=u_{1}{\rm e}^{i\phi_{1}}{\rm
e}^{i\alpha_{1}L_{1}r\frac{P}{Q}}T_{x}^{r}+u_{2}{\rm e}^{i\phi_{2}}
  {\rm e}^{i\alpha_{2}L_{2}r\frac{P}{Q}}T_{y}^{r}+ c.c.
  \end{equation}
  or  can be simply rewritten  as:
 \begin{equation}
  U_{0}=u_{1}\alpha T_{x}^{r}+v_{1}\beta
T_{y}^{r}+u_{1}\alpha^{-1}T_{x}^{-r}+v_{1}\beta^{-1}T_{y}^{-r}
\end{equation}
where $\alpha$ and $\beta$ are defined by
\begin{eqnarray}
  \alpha&=&{\rm e}^{i\phi_{1}}{\rm e}^{i\alpha_{1}L_{1}r\frac{P}{Q}}
\nonumber\\
\beta&=&{\rm e}^{i\phi_{2}}
{\rm e}^{i\alpha_{2}L_{2}r\frac{P}{Q}}
\end{eqnarray}
Taking
\begin{eqnarray}
T_{x}^{r}=T^{r}(\frac{\vec{L_{1}}}{N_{s}})=T(\frac{r\vec{L}_{1}}{N_{s}})=
\tilde{T}_{x}   \nonumber\\
T_{y}^{r}=T^{r}(\frac{\vec{L_{2}}}{N_{s}})=T(\frac{r\vec{L}_{2}}{N_{s}})=
\tilde{T}_{y}
\end{eqnarray}
into account and noting the  fact that $\tilde{T}_{\pm{x}}$ and
$\tilde{T}_{\pm{y}}$ form
 another  algebra   similar to that of
$T_{\pm{x}}$ and $T_{\pm{y}}$,  without loss of  generality, we obtain
the following form for impurity potential $U_{0}$:
\begin{equation}
U_{0}=u_{1}\alpha T_{x}+v_{1}\beta
T_{y}+u_{1}\alpha^{-1}T_{x}^{-1}+v_{1}\beta^{-1}T_{y}^{-1}
\end{equation}

Making use of  the perturbation theory for  degenerate case, we have the
secular
equation:
\begin{equation}
U_{0}\Psi_{n,k}=\varepsilon\Psi_{n,k}
\end{equation}
where $\varepsilon$ is the first order correction of  energy and $\Psi_{n,k}$
is the degenerate wave function.

A more general case has been discussed by L. D. Faddeev and R. M. Kashaev [11].
 Eq.(108)
is nothing but a special case of Ref.[11] by setting $\rho=0$ in their paper.
 Therefore, we can directly quote their results:
\begin{equation}
\varepsilon=-u_{1}u_{2}(q-q^{-1})\sum_{m=1}^{Q-1}z_{m}+(u_{1}+u_{2})
(q^{\frac{1}{2}}+q^{-\frac{1}{2}})
\end{equation}
where $z_{m}$ can be determined by the generalized Bethe-Ansatz Equation:
\begin{equation}
q^{-\frac{1}{2}}\frac{(u_{1}z_{l}+\frac{1}{2})(u_{2}z_{l}+\frac{1}{2})}
{(q^{\frac{1}{2}}u_{2}z_{l}-1)
(q^{\frac{1}{2}}u_{1}z_{l}-1)}
=\prod_{m=1,\ne l}^{Q-1}\frac{qz_{l}-z_{m}}{z_{l}-z_{m}}
\end{equation}
where $l=1,2,...,Q-1$.
The integer number $Q$ is just the number appearing in $\nu=\frac{P}{Q}$.

{}From the above discussion we can see that the energy is splitted into many
subbands
in the presence of weak impurity potential and can be described in terms of the
theory
presented in Ref.[11].  This is the main result of this paper.

\section{Concluding Remarks}
In the above discussions we have pointed out that the degeneracy for FQHE can
be
 described in terms of  the cyclic representation of quantum algebra. As is
shown
 in Ref.[7], when $q^{Q}=1$, the dimension of irreducible cyclic representation
  of quantum algebra associated with $sl_{q}(2)$ should be $Q$. Therefore our
conclusion
   is consistent with the  general results of De Concini and Kac [7]. The
quantum
   algebra is introduced based on the relations  Eq.(72) and
    Eq.(73), which is nothing
   but the quantum plane  defined in Ref. [27]. Therefore, the degeneracy
properties
    can be read from  the geometry on the quantum plane and described
    in terms of the $q$-boson operators shown by Eq.(18)-(20). The
determination of
    dimension is complicated and  some of  discussions had been  made in
Refs.[6,7,20]. In this
    paper we have discussed the weak impurity, namely, only leading term of
    the potential $U$ is survived,
 which removes the degeneracy
    for FQHE.  Fortunately,
     the leading term $U_{0}$ can be expressed through $T^{r}_{\pm x}$ and
$T^{r}_{\pm y}$.
    We then are able to quote all the results in Ref.[11]. It is also
attractive to calculate
     the overlaping  between  the ground state and the first excited state due
to
     strong impurity. To do this, it may be beyond the present quantum
algebraic
     structure. A new approach including more complicated algebra may be
deserved.

\vspace{0.5cm}
\begin{flushleft} \Large \bf
Acknowledgements
\end{flushleft}
The authors acknowledge Prof. Y. S. Wu, Prof. L. Yu and Drs  F. Caitan, H. C.
Fu,
L. M. Kuang for valuable discussions, They also thank Miss Y. J. Wu for reading
 and typing the paper. This research is partly  supported by the National
Natural
  Science Foundation of China.

\end{document}